# Improving Students Performance in Small-Scale Online Courses

## A Machine Learning-Based Intervention

Sepinoud Azimi[(✉)]
Åbo Akademi University, Turku, Finland
sepinoud.azimi@abo.fi

Carmen-Gabriela Popa
University Politehnica of Bucharest, Bucharest, Romania

Tatjana Cucić
University of Novi Sad, Novi Bečej, Serbia

**Abstract**— The birth of massive open online courses (MOOCs) has had an undeniable effect on how teaching is being delivered. It seems that traditional in-class teaching is becoming less popular with the young generation – the generation that wants to choose when, where and at what pace they are learning. As such, many universities are moving towards taking their courses, at least partially, online. However, online courses, although very appealing to the younger generation of learners, come at a cost. For example, the dropout rate of such courses are higher than that of more traditional ones, and the reduced in-person interaction with the teachers results in less timely guidance and intervention from the educators. Machine learning (ML)-based approaches have shown phenomenal successes in other domains. The existing stigma that applying ML-based techniques requires a large amount of data seems to be a bottleneck when dealing with small-scale courses with limited amounts of produced data. In this study, we show not only that the da-ta collected from an online learning management system could be well utilized in order to predict students' overall performance but also that it could be used to propose timely intervention strategies to boost the students' performance level. The results of this study indicate that effective intervention strategies could be suggested as early as the middle of the course to change the course of students' progress for the better. We also present an assistive pedagogical tool based on the outcome of this study, to assist in identifying challenging students and in suggesting early intervention strategies.

## 1 Introduction

With the unprecedented technological advancement over the past decade, interest in pursuing online courses as well as online degrees has increased sharply. Participating in a top-rated university class or obtaining a degree from a renowned institution is no longer a dream for the majority. No matter the geographical distance or the financial barriers, thanks to massive open online courses (MOOCs), high-quality education is now a dream come true for many learners. This liberation of education has forced universities to reconsider their modes of teaching. To compete with the online overseas counterparts, and to appeal to the younger generation of learners, many universities now tend to offer their courses online. Moreover, online education is expected to go mainstream by 2025; see [1] and as such the research on the factors influencing students' performance is gaining more attention, see [2,3]. However, this mode of education has its own drawbacks. Student dropouts, lack of personal interaction and teachers' limited familiarity with the students are just some such hurdles. The majority of university online courses, with their one-size-fits-all approach, require students to flourish in their studies without any individual guidance from the educators. Such approach, although instrumental to scaling up the courses, could lead to a low performance level in the students and high dropout rates.

Much research has been applied to finding ways to reduce the high dropout rate with MOOCs; see for example [4, 5, 6]. Researchers have also investigated approaches to predicting students' performance in MOOCs and proposing proactive interventions - see [7, 8, 9] - using machine learning (ML) models. However, such attempts have targeted large MOOCs with an abundant pool of data to benefit from. It remains challenging to design similar tools to assist educators involved in much smaller scale university online courses. Such data-hungry ML-based approaches fail to perform as well at university course level. However, a smaller sized course does not necessarily mean a lack of useful data. Small-scale university courses produce data that is typically overlooked and underutilized. The learning management systems (e.g. Moodle) that are the hosts to the majority of university course regularly gather information on student behaviour, in terms of how the students interact with the online system, as well as the records of their graded assignments. Moodle (https://moodle.org) is one of the major learning management systems; it collects data on student learning activities – from watching videos, to participating in forum discussions, to attempts at quizzes, to submitting assignments. In short, this data projects how a student has interacted with the course, how much time s/he has invested in learning the content and how s/he interacted with other peers. To some extent, such data mirrors how a student would have behaved in an in-class course.

In this paper we propose an approach to identifying the underlying pattern in students' behaviors that leads to different overall performances. Such an approach would provide effective pedagogic guidance and timely intervention strategies to assist each student in improving their individual style of learning. It would also, to some extent, compensate for the lack of educators' individual familiarity with the students. In this paper we aim to answer the following research questions:

RQ1. Could student's activity logs provide insight for diagnosing students' low performance early on?

RQ2. When could intervention strategies be applied in order to boost student's overall performance?

RQ3. What early intervention strategies could be used in order to boost student's overall performance?

This paper contributes to the body of literature by proposing an ML-based approach, utilizing the data of small-scale university online courses, in order to provide pedagogical intervention strategies during the first half of the course. It is important to place the intervention as early as possible in the course to diagnose and prevent any drop in the students' overall performance.

The paper is structured as follows. In Section 2, we present a short overview of the existing literature on the topic of this study. In Section 3, we describe the data processing and modelling steps of this study. In Section 4, we present the results of our analysis. In Section 5, we discuss the implications of our results and present an educational tool we developed based on the findings of this study. Finally, we present some concluding remarks in Section 6.

## 2   Background and State of the Art

Online education has gained popularity over the past few years and the ever-growing number of online courses has caused a dramatic change in the teacher's traditional roles. In today's educational system, rather than being a mentor and an advisor, teachers are taking up the role of facilitators. Such an approach along with the students' freedom to choose the time, the place and the pace of the study, if designed carefully, could benefit the more independent-thinking students and enhance course enrolment statistics. However, the educational system should deal with these advancements carefully, as they could lead to a neglecting of the students who are not necessarily autonomous and need personal advice to perform well in their studies. There is a growing body of literature focusing on predicting students' performance in MOOCs; see for example [4-9].

However, the research on smaller scale courses is lagging behind. To the best of our knowledge, even this small number of existing studies does not begin to solve the problem of low-performing students from all directions. Such studies either focus on predicting the students' overall performance without proposing any intervention strategies or they use simple statistical visualization-based approaches to data analysis. The studies using ML-based approaches either only classify the final overall grades of the students or do not attain a high level of accuracy, or they fail to meet the predefined goals. In what follows we map out some of the existing literature in the field.

In [10], the authors looked into the students' behavior with regard to delivering their home assignments for an online physics course. They used the k-nearest neighbor approach and genetic algorithm on the data collected from 227 students to cluster the students and predict their final grades. In a similar study, [11], the authors used recommender systems to predict the performance of students from a specified course. They used a publicly available dataset from the Knowledge Discovery and Data Mining (KDD) Challenge 2010. The main features were knowledge components (KCs) and opportunity counts (OCs). KCs were specific skills used for solving the problem, while OCs showed the number of times that specific KC had been encountered previously.

Another recommender-based approach was presented in [12], where the authors used multiregression and matrix factorization algorithms to classify students and predict the dropout rate.

Since Moodle is widely utilized as one of the major learning management systems, several studies based their research on the data collected from Moodle. In [13], the authors developed a tool to identify students at risk of failing and to evaluate the results of pedagogic measures taken to prevent the failure of a student. This paper mainly used the visualization of data collected from Moodle logs as the basis of its analysis. Similar to the previous study, there have been several attempts at using simple statistical tools to analyze Moodle log data. One such tool is the learning analytic tool available in Moodle. For example, [14] uses the Moodle learning analytic tool on the data collected from 337 students over three years to identify the variables correlated with the final grades. The results indicated that failing the course correlated with the student's negative attitude towards Moodle. Also, good grades were associated with more frequent use of the learning platform. Another study that used statistical analysis tools is presented in [15]. The authors used ANOVA, the Chi-square test and linear regression to explore the relation between students' online interactions and students' grades. The data used in the paper consisted of log-in frequency such as the last log-in time and time spent online, and the main research question was to determine the extent of reliability of individual variables derived from log data as predictors of academic success.

Machine learning techniques have also been among the popular approaches for Moodle data analysis over the past decade. In one of the early studies, in [16], the authors used various machine learning techniques to predict overall performance using mainly students' prior assignment grades to predict their overall performance.

The relatively small amount of data, extracted from university-based online courses, is one of the major challenges of using machine learning approaches. This lack of data could potentially lead to a low accuracy in the developed models. For example, in [17], the authors used a decision tree–based model to determine the learning style of 35 students enrolled on a Moodle-based course. Receiver operating characteristic (ROC) curve was used to assess the quality of the results. The data was divided into several training and testing sets with an average accuracy of 76%. In another study, [18], the authors used backward step-wise multiple regression analysis to determine whether there was a connection between students' Moodle logs and their academic success. However, the results indicated no correlation. The authors stated that the small dataset could be the reason behind such an outcome.

The literature on students' performance in online courses has attempted either to predict the final outcome or to identify the students' learning style. The limited number of studies that discussed intervention strategies do not propose any approach to implementing such an intervention as early as possible. The limited amount of data in university-based online courses is another obstacle that needs special attention. Using only assignment grades, as performed for example in [18], may not be enough to obtain a reliable outcome. Moreover, this approach does not provide any insights into understanding the students' personal challenges. In this study, we propose to use both machine learning techniques and in-person coaching as a means to gaining optimal results. We also use the data related to the students' interactions with a learning management system as well as their assignment grades. Rather than focusing on the data from the entire duration of a course, we shift our attention to the first half of the course so that any proposed intervention strategies could be applied in a timely manner in order to change the students' overall performance trajectory for the better. The rest of the data is utilized in the form of an assistive tool for supervising students' progress during the second half of the course.

## 3    Methodology

In this section we apply the methodologies used in this study – established from the data collected – to developing a model. The study's workflow is depicted in Figure 1.

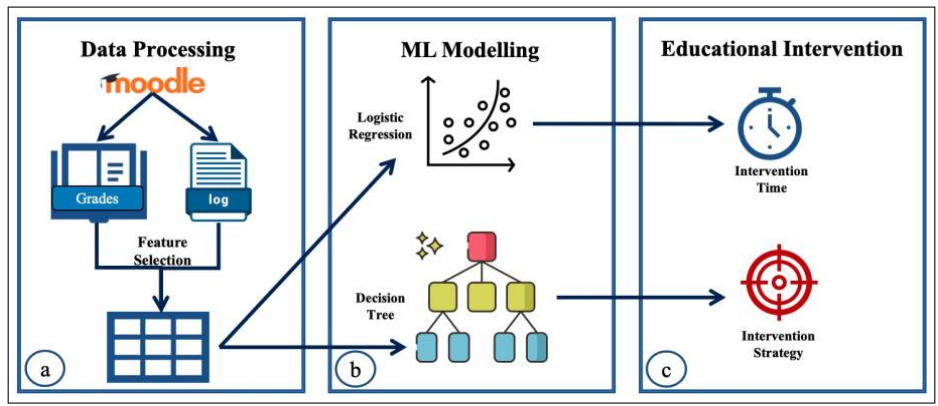

**Fig 1. Study's workflow**: (a) Students' assignment grades as well as their interactions log files are collected from Moodle platform; the data is further processed and the relevant features are selected, (b) a logistic regression and a two decision tree models are trained based on the collected data, (c) the outcome of the modelling in the previous step facilitates deciding on the right time and right strategy to intervene in order to improve students' overall performance

### 3.1    Data Collection

The data was collected from a fully online nine-week-long course on machine learning at a Finnish university, hosted on the online learning management system Moodle. The dataset contained anonymized information relating to 107 enrolled students. The data included students' grades (from 3 mini projects, 3 quizzes and 3 peer reviews and the final overall grade) as well as the course logs. The deadline for the three mini projects fell within weeks 3, 5 and 8 of the course, whereas the deadline for the quizzes fell within weeks 2, 4 and 8. The students' final grade is ranked between 0 and 5, where 0 indicates they have failed the course. The students are classified into the following groups: dropouts (grade 0), low achievers (grades 1–3) and high achievers (grades 4 and 5). It is worthy of note that in the current dataset, no student obtained a grade of 1 (only just a pass). Figure 2 depicts the distribution of students according to their final grade. Of those who failed the course, 27% had submitted some assignments, while only some interactions were recorded for the remainder.

The course logs constitute the information collected throughout the duration of the course in relation to students' interaction on the course online platform. Moodle logs several different activities. In the course under study, there were 20 different types of activities registered in Moodle, which were categorized as follows:

- course content related (course module viewed, course viewed, course activity completion updated, course module instance list viewed, content page viewed, lesson started, lesson resumed, lesson restarted, lesson ended)
- assignment related (quiz attempt reviewed, quiz attempt submitted, quiz attempt summary viewed, quiz attempt viewed, quiz attempt started, question answered, question viewed, submission re-assessed, submission reassessed, submission updated, submission created, submission viewed)
- grade related (grade user report viewed, grade overview report viewed, user graded, grade deleted, user profile viewed, recent activity viewed, user report viewed, course user report viewed, outline report viewed)
- forum related (post updated, post created, discussion created, some content has been posted, discussion viewed)

It is important to note that, in this data set, the students are not equally distributed among the grade classes. To balance the dataset, we used the synthetic minority over-sampling technique (SMOTE); see [19]. The final dataset was then divided into training (80% of the data) and testing (20% of the data) datasets.

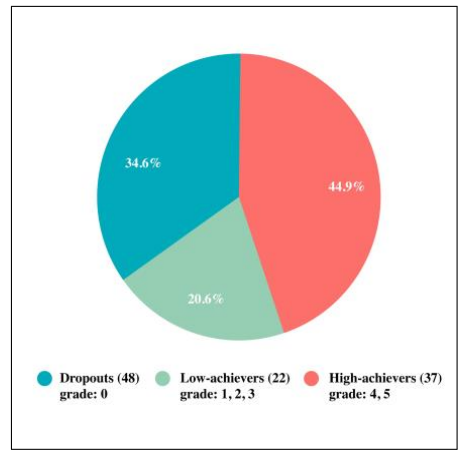

**Fig 2.** Grade Distribution.

### 3.2 Modelling

In this section we present the ML modelling approaches used in this study, i.e. logistic regression and decision trees. We have used two very simple and comprehensible approaches that are suitable for a small-scale dataset for a not very complex problem. Both models are highly explainable and interpretable, which makes them an ideal foundation for designing intervention strategies

Logistic Regression One of the most widely used machine learning approaches for prediction and diagnosis is logistic regression (LR). LR models are very effective for solving less complex problems involving small datasets; see [20]. LR approaches are not computational-resources-demanding, and they are highly interpretable. LR is a regression method for predicting a categorical-dependent variable by using the maximum-likelihood ratio; see [21]. In an LR model, the categorical-dependent variable could have two or more levels. Unlike the dependent variable, the independent variables could be both numerical and categorical; see for example [22].

In this study we trained LR models on the training dataset using the Scikit-learn's L-BFGS solver, [23], which is suitable for datasets with multinomial classes. In order to answer our first and second research questions, (RQ1, RQ2), we trained three different RL models on the subsets of the training dataset – i.e. the data including only the grades, the data including only the logs, and the data including both the grades and the logs. The detailed results of the analysis are presented in Section 4.

Decision Trees "Knowledge discovery in databases" (KDD) or data mining (DM) is the process of untangling the web of unstructured data and identifying meaningful causal relationships, [24]. DM approaches are instrumental when it comes to making the right decision; see [25]. One of such DM approaches is decision tree. Decision tree (DT) models are trained by recursively partitioning the data space and fitting a simple prediction model within each partition; see [26]. The final model could be represented as a tree shaped structure which facilitates both classification and prediction; see [27]. Unlike many other machine learning algorithms, DTs require minimal data processing and they perform well even with a small-scale dataset. Their tree-shaped representation makes them very intuitive and easy to interpret and explain. Several algorithms have been proposed to train a DT, for example: ID3 (Iterative Dichotomizer 3), [28]; C5.0, [29], classification and regression trees (CART); [30]; and chi-squared automatic interactive detector (CHAID), [31].

We have chosen to use the CART algorithm in this study as it is ideal for multi-class classification, is suitable for capturing non-linear relationships between influential factors, and provides the highest degree of model interpretability. To answer the third research question, (RQ3), the results of the LR analysis as well as the answer to (RQ2) were used. We trained a CART decision tree model using data from the first five weeks in order to identify causal relationships as the basis for devising intervention strategies. The detailed results of the analysis are presented in Section 4.

## 4   Results

In this section we provide the detailed results of our preliminary data analysis and ML modelling.

### 4.1 Learning Platform Data Analysis

A preliminary data analysis based on the students' interactions with the course Moodle page revealed interesting results. In this study, an interaction is defined simply as a student's mouse click within the online platform. Based on the interaction data analysis, as expected the students' interactions correlated with their final grade – i.e. the students who failed the course had a small number of interactions, with an average of 92 interactions per student; low achievers had an average of 273 interactions per student; and high achievers had a considerably larger number of interactions, with an average of 450 per student. An interesting point to note is that, among the high achievers, the students with a final grade of 4 had significantly more interactions than those with a final grade of 5 (maximum); see Figure 3.

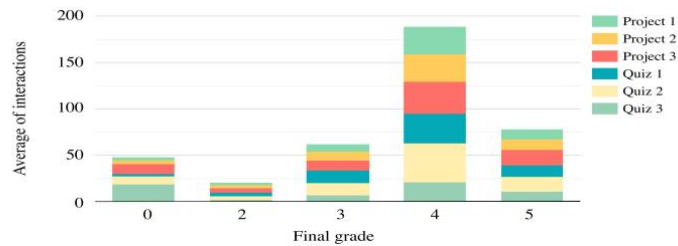

**Fig 3.** Average of Interaction for each assignment type per grade.

This interpretation could indicate that the students with an overall grade 4 could benefit from one-to-one intervention sessions. Another interesting point to note is that, between weeks 4 and 6 (halfway through the course), the students had the largest number of interactions; see Figure 4. This could indicate that an intervention should perhaps be introduced before week 6, when the students are highly engaged with the course. It is also worth noting that the fourth and sixth week of the course corresponds to the weeks with the deadline for quiz assignments. This observation could indicate that, for quizzes, students rely more on course content, whereas for their mini projects they most likely search for information and guidelines online.

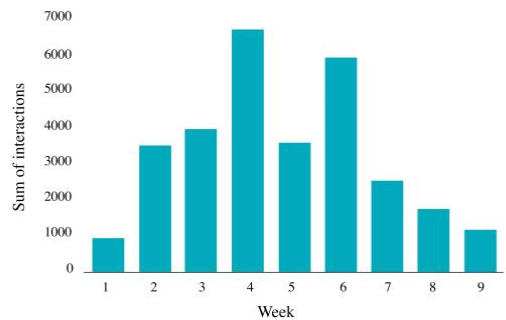

**Fig 4.** Average Interaction Per Week.

### 4.2 Early Performance Prediction

As mentioned in Section 3.2, we trained several LR models to answer (RQ1) and (RQ2). We specifically wanted to know (1) whether including the log file information in the dataset would bring about a better predictive outcome, and (2) whether we could train an accurate LR model with the data from the first few weeks of the course to predict the students' overall performance. Since one of the main objectives of this paper is to provide guidance on early interventions, we started training the LR models based only on the first week's data and we, then, gradually included the data from the subsequent weeks. The results of our experiment are shown in Table 1. As it can be observed from the table, the models trained on Grades + Logs data consistently resulted in a higher degree of accuracy than with the models trained only on the grades or on the logs. It could also be observed that including the data of the log files led to models with an acceptable level of accuracy using the data available during the first few weeks of the course. As such, answering to (RQ1), the results indicate that including the log information would bring insightful information as early as week 3 in the course. The result of this experiment also provides answer to (RQ2). It can be seen from Table 1 that there is a boost in the models accuracy from week 3 and the accuracy reaches an acceptable level in week 5. Based on the outcome of this step we trained our DT model on the training dataset including the data from the first five weeks of the course.

| Weeks | Grades. | Logs | Grades + Logs |
|---|---|---|---|
| 1 | 0%* | 19% | 19% |
| 1 - 2 | 30% | 29% | 47% |
| 1 - 3 | 53% | 42% | 74% |
| 1 - 4 | 57% | 53% | 79% |
| 1 - 5 | 71% | 52% | 84% |
| 1 - 6 | 71% | 55% | 85% |
| 1 - 7 | 81% | 51% | 89% |
| 1 - 8 | 82% | 58% | 92% |
| 1 - 9 | 82% | 66% | 94% |

**Table 1.** Prediction accuracy based on limited data. *No grade available for this case.

### 4.3 Early Intervention

To identify the causal relationships as the basis for proposing intervention strategies and thus answering (RQ3) data from the first 5 weeks of the course was used to train a DT model. Figure 5 presents the final DT model. In the DT representation, "Stat0" stands for the interaction with the course content, "Stat1" stands for the interaction with the assignments, "Weekn" stands for week n of the course, "MPn" stands for the grade of mini project n and "Quizn" stands for the grade of quiz n. The final model had 96% accuracy on training and 91% accuracy on testing dataset. The paths involving the majority of the cases are highlighted in Figure 5; the other paths only included fewer than three cases and were thus not included in the analysis. Table 2 shows the DT model's precision and recall breakdown based on the final grades. As it can be seen from Table 2, the DT model performs well in terms of both precision and recall for all classes. The analysis of the DT model reveals interesting results. The mini project on week 5 has a decisive impact on differentiating between low achievers and high achievers – i.e. the students with 83% or more of the project's maximum grade ended up with an overall grade of 4 or 5. This means that one of the main focuses of the intervention strategies could be on, for example, providing more coaching for this specific mini project. This is an interesting observation as this is the second mini project out of three and it is placed early enough in the course to give the students an opportunity to recover from any mishaps they may have encountered early in the course.

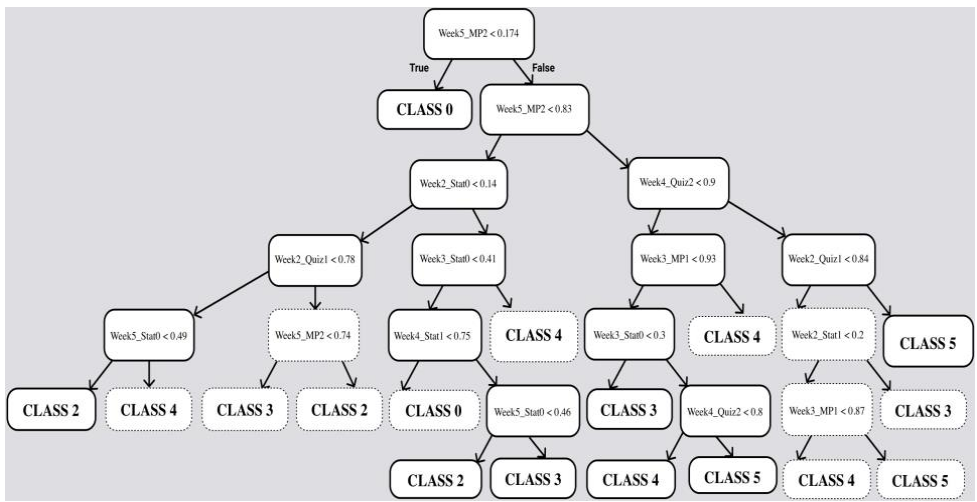

**Fig 5.** Decision tree model based on the data from the first 5 weeks of the course.

|  | Training | | Testing | |
|---|---|---|---|---|
| Grade | Precision | Recall | Precision | Recall |
| 0 | 0.97 | 0.99 | 1.00 | 1.00 |
| 2 | 0.98 | 0.98 | 0.88 | 1.00 |
| 3 | 0.97 | 0.98 | 0.92 | 0.85 |
| 4 | 0.93 | 0.93 | 0.86 | 0.85 |
| 5 | 0.97 | 0.89 | 1.00 | 0.95 |

**Table 2.** Precision and recall breakdown based on the final grades

The DT model shows that for those students who have not interacted with the course content during the early weeks of the course (Week3 Stat0 < 0.41), the time they invest into studying the material of week 5 could be a decisive factor in achieving an overall grade of 3 rather than 2 – i.e. the students with 46% or more of the maximum number of interactions (Week5 Stat0) scoring an overall grade of 3 rather than 2. The DT model also suggests that, for those students who have not performed well on their first mini project in week 3 (Week3 MP1 < 0.93), the attention on the week 3 course content (Week3 Stat0) could be a factor in predicting whether they are assigned to the class of low achievers or high achievers; similarly, the same path suggests that even for students on this path, focus strongly on the week 4 assignment (Week4 Quiz2) could place them in the class of the students with the maximum overall grade.

An important take-away message from studying the DT model is that a low performance in the first half of the course does not need to be a determinant of an overall low performance of a student and that a timely change in attitude could lead to being in the group of high achievers.

We also used the data from the second half of the course to train a second DT model with 98% accuracy on training and 93% accuracy on testing dataset. This second model includes information on students' possible trajectories towards different overall grades using the data from week 5 onwards. As such, the DT model could be used as a supervision tool, see Figure 6. After the initial early intervention, the teacher could observe if the students are following the path to the desired class of overall grade or not. If needed, consequent in-person meetings could be arranged with the students to guide them towards the right direction. As an example, let us consider a student who had graded a low grade for the second project in week 5 (Week5 MP2 < 0.67) and in the one-to-one intervention session it is decided that s/he has a chance of still ending up in the class of high achievers. However, the grade for the third project in week 7 is not as high as expected (0.6 ≤ Week7 MP2 < 0.85). In a second one-to-one session the teacher could advise the student to focus more on the learning content of Week 8 (Week8 Stat0 ≥ 0.62).

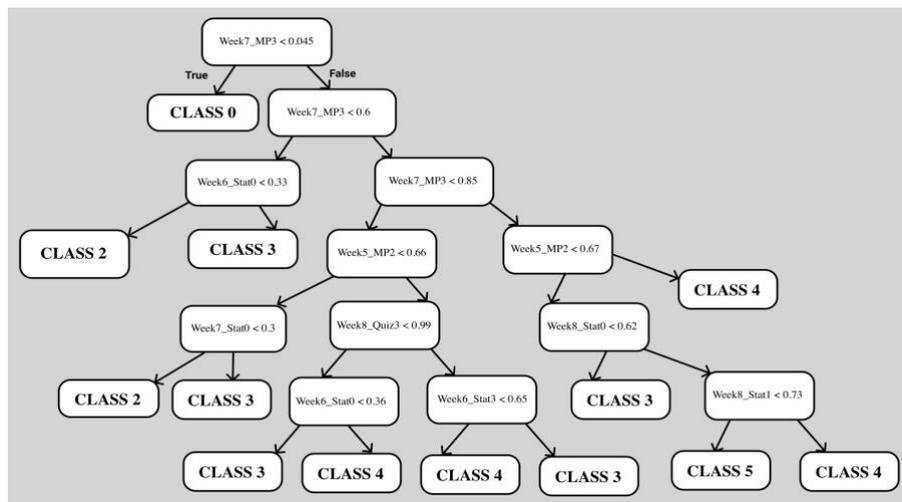

**Fig 6.** Decision tree model based on the data from weeks 5 to 8.

## 5 Discussion

One of the common features of most studies that focus on students' performance in online courses is the utilization of data collected throughout the entire duration of the courses under study. Although this approach benefits from a larger pool of data, it fails to provide a meaningful action plan early on, to prevent poor student performance. In reality, the fate of a student is decided, based on her/his experience in the course, much earlier. As discussed in Section 4, it is still possible to identify the right time and right strategy to intervene in order to improve students' overall performance, using only a subset of the data available. Our results indicated that reliable strategies could be proposed as early as week 4 (for a nine-week course), in order to change the course of students' progress. This approach, however, does not discard the rest of the data. As detailed in Section 4, the remaining data (up to one week before the end of the course) could be used as a powerful supervision tool to ensure students stay on the path that leads to their best possible outcome. One should remember that although the approach proposed in this paper (as well as any other solely machine-based approaches) is a powerful tool for identifying struggling students, it could not replace the personal insight of the teacher. It still remains the teacher's responsibility to learn about his/her students' individual needs, which could differ dramatically for each student. The approach presented here is an assistive tool to identify those students who need in-person attention in classes with large numbers of students and limited human resources. Using the outcome of this study, we have developed a new tool, OLIT (Online Learning Improvement Tool). As the input, OLIT reads an input file that includes the data from the first five weeks of the course (log files + grades). Then, it predicts the final overall grade of each student in the dataset. After this step, the teacher could select an individual student from a list and read her/his individual improvement strategy. Figure 7 presents a snapshot of the tool, the relevant code and a small sample file accessible on request. As Moodle is a free and open-source learning system, the developed tool of this study could be integrated as a plug-in into Moodle. The tool could automatically alert the teacher when a student is in danger of an overall low performance or even dropping out of the course. Based on the information provided by the tool, the teacher could then arrange a face-to-face meeting with the student and mutually agree on the follow-up steps.

## 6 Discussion

Online education is a welcomed consequence of the new era of technological advancements. It facilities inclusion and diversity, brings enormous flexibility regarding when and where to study, and provides access to high-quality education. It also leads to a loss of the personal touch in most cases. Excluding this human touch from the educational cycle has its own repercussions – specifically, less timely guidance and intervention from the educators. In this study, we proposed an approach to bring together the benefits of both worlds (modern online and traditional in-class educational system). We presented an approach which uses the students' assignment grades as well as their log file information to train several ML-based models. The ML models facilitate decisions about the right time and right strategy to intervene in order to improve students' overall performance in an online course. We then applied our findings to the development of an assistive online tool. The contribution of this study is multifold.

Unlike the other existing literature, this study proposes an approach that provides intervention strategies as early as the first half of the course period. This is an important result as waiting too long to guide the students into the right path could lead to discouragement and loss of motivation. Another added value is that such an approach involves utilizing the data from the second half of the course as the basis to supervise the students' progress following an intervention session. We also believe the approach proposed here is useful not only in the context of online courses but also as it could be further developed to address students' overall performance throughout the entire duration of their university studies. This paper also rejects the common belief that a small-scale course must inevitably mean a lack of useful data. In this study, we showed that the combination of the students' interaction data as well as their graded assignment records is sufficient to be used as the basis for ML-based models with a good level of accuracy. This means that to take advantage of any existing smart ML-based solution, a teacher need not wait for several versions of a course to be completed – even just one could provide meaningful insight.

This study, inevitably, has some limitations. The approach and the tool proposed in this paper are solely designed for courses with a similar structure to that used as the data source here. However, we believe the approach, with some modifications, could indeed be applied to courses with different structures. The dataset is also relatively small (only one version of one course) and it is not inclusive (it misses one class of grades).
With further data collection from future iterations of the course and fine-tuning of the models used in this study, this limitation could be overcome.

For our future work, we plan to integrate the tool in our own university's Moodle system and test it during the next iterations of the course. We plan to collect the data from four upcoming courses of similar structure during the next academic year. We aim to employ the deployed tool in two of those courses (test group) and apply intervention strategies for their students. Finally, we plan to compare the overall performance of the students in the test group with that of students in the control group. We will also use the data from all four courses to fine-tune the existing models and train new ML models which could identify more complex causal relations.

## 8 Authors

**Sepinoud Azimi** is an Adjunct Professor in Biomedical Data Analytics at the Department of Information Technologies at Åbo Akademi University, Turku, Finland.

**Carme-Gabriella Popa** is a Computer Science student at the Faculty of Automatic Control and Computing, University Politehnica of Bucharest, Bucharest, Romania.

**Tatjana Cucić** is a graduate student at the Faculty of Sciences, University of Novi Sad, Serbia.